\documentclass[11pt,a4paper]{article}
\usepackage{graphicx}
\usepackage{amsmath}
\usepackage{amssymb}
\usepackage{amsthm}
\usepackage{subfigure}
\usepackage{cite}

\theoremstyle{plain}

\providecommand{\keywords}[1]
{\textbf{Keywords:} #1}
	
\newcommand{\bm}{\boldsymbol}
\newcommand{\vect}[1]{\bm{#1}}
\newcommand{\mat}[1]{\bm{#1}}

\newcommand{\E}[1]{{\mathbb{E}}\left\{#1\right\}}

\newcommand{\etr}[1]{{\mathrm{etr}}\{#1\}}

\renewcommand{\det}[1]{\left|#1\right|}

\newcommand{\vone}{\vect{1}}
\newcommand{\vt}{\vect{t}}
\newcommand{\vx}{\vect{x}}

\newcommand{\vy}{\vect{y}}
\newcommand{\vytilde}{\tilde{\vy}}
\newcommand{\vz}{\vect{z}}
\newcommand{\vzbar}{\bar{\vz}}

\newcommand{\I}{\mat{I}}
\newcommand{\M}{\mat{M}}

\newcommand{\Porth}[1]{\mat{P}^{\perp}_{#1}}
\newcommand{\Q}{\mat{Q}}
\renewcommand{\S}{\mat{S}}

\newcommand{\Z}{\mat{Z}}

\newcommand{\vmu}{\bm{\mu}}
\newcommand{\mSigma}{\bm{\Sigma}}
\newcommand{\invSigma}{\mSigma^{-1}}

\newcommand{\xZ}{\begin{bmatrix} \vx & \Z \end{bmatrix}}
\newcommand{\xZtranspose}{\begin{bmatrix} \vx & \Z \end{bmatrix}^{T}}

\newcommand{\cxZ}{\begin{bmatrix} \vx - \vmu & \Z - \vmu \vone_{n}^{T} \end{bmatrix}}
\newcommand{\cxZtranspose}{\begin{bmatrix} \vx - \vmu & \Z - \vmu \vone_{n}^{T} \end{bmatrix}^{T}}

\newcommand{\yZ}{\begin{bmatrix} \vy & \Z \end{bmatrix}}
\newcommand{\yZtranspose}{\begin{bmatrix} \vy & \Z \end{bmatrix}^{T}}

\newcommand{\ytildeZ}{\begin{bmatrix} \vytilde & \Z \end{bmatrix}}

\newcommand{\cyZ}{\begin{bmatrix} \vy - \vmu & \Z - \vmu \vone_{n}^{T} \end{bmatrix}}
\newcommand{\cyZtranspose}{\begin{bmatrix} \vy - \vmu & \Z - \vmu \vone_{n}^{T} \end{bmatrix}^{T}}

\newcommand{\cytildeZ}{\begin{bmatrix} \vytilde - \vmu & \Z - \vmu \vone_{n}^{T} \end{bmatrix}}
\newcommand{\cytildeZtranspose}{\begin{bmatrix} \vytilde - \vmu & \Z - \vmu \vone_{n}^{T} \end{bmatrix}^{T}}

\newcommand{\meanyZ}{\begin{bmatrix}  \vmu & \vmu \vone_{n}^{T} \end{bmatrix}}
\newcommand{\meantyZ}{\begin{bmatrix}  \alpha \vt + \vmu & \vmu \vone_{n}^{T} \end{bmatrix}}

\newcommand{\GLR}{\mathrm{GLR}}
\newcommand{\GLRam}{\mathrm{GLR}_{\text{\tiny{AM}}}}
\newcommand{\GLRrm}{\mathrm{GLR}_{\text{\tiny{RM}}}}
\newcommand{\GLRmm}{\mathrm{GLR}_{\text{\tiny{MM}}}}

\newcommand{\Pfa}{P_{fa}}
\newcommand{\Pd}{P_{d}}

\newcommand{\matrixN}[5]{\mathcal{N}_{#1,#2}\left(#3,#4 \otimes #5\right)}
\newcommand{\matrixT}[6]{\mathcal{T}_{#1,#2}\left(#3,#4,#5,#6\right)}

\newcommand{\dist}{\overset{d}{=}}

\begin{document}
\title{Sub-pixel detection in hyperspectral imaging with elliptically contoured $t$-distributed background}
\author{ Olivier Besson and Fran\c{c}ois Vincent\thanks{The authors are with University of Toulouse, ISAE-SUPAERO, Toulouse, France. Email:  olivier.besson@isae-supaero.fr,  francois.vincent@isae-supaero.fr.}}
\date{March 2020}
\maketitle	
\begin{abstract}
Detection of a target with known spectral signature when this target may occupy only a fraction of the pixel is an important issue in hyperspectral imaging. We recently derived the generalized likelihood ratio test (GLRT) for such sub-pixel targets, either for the so-called replacement model where the presence of a target induces a decrease of the background power, due to the sum of abundances equal to one, or for a mixed model which alleviates some of the limitations of the replacement model. In both cases, the background was assumed to be Gaussian distributed. The aim of this short communication is to extend these detectors to the broader class of elliptically contoured distributions, more precisely matrix-variate $t$-distributions with unknown mean and covariance matrix. We show that the generalized likelihood ratio tests in the $t$-distributed case coincide with their Gaussian counterparts, which confers the latter an increased generality for application. The performance as well as the robustness of these detectors are evaluated through numerical simulations.
\end{abstract}
\keywords{Detection, generalized likelihood ratio test, hyperspectral imaging, replacement model, Student distribution.}	
\section{Problem statement}
Hyperspectral imaging has become an increasingly popular tool for remote sensing and scene information retrieval, whether for civil or military needs and in a large number of applications, including analysis of the spectral content of soils, vegetation or minerals, detection of man-made materials or vehicles, to name a few \cite{Eismann12,Manolakis16}. One of the challenges of hyperspectral imaging is to detect a target -whose spectral signature is assumed to be known- within a background whose statistical properties are not fully known \cite{Manolakis02,Manolakis14,Nasrabadi14}. Depending on the spatial resolution of hyperspectral sensors and the size of the target, the latter may occupy the totality or only a fraction of the pixel under test (PUT), in which case one speaks of sub-pixel targets. In the latter case, the target  replaces part of the background in the PUT, leading to the so-called replacement model \cite{Manolakis01b,Manolakis04}. 

Whatever the case, full-pixel or sub-pixel targets, the problem can be formulated as a conventional composite hypothesis problem \cite{Manolakis02,Manolakis14,Nasrabadi14,Manolakis01b,Manolakis04,DiPietro10,Frontera17}: given a vector $\vy \in \mathbb{R}^{p}$ -where $p$ denotes the number of spectral bands used- which represents the reflectance in the PUT, is there a component along $\vt$ -the signature of interest (SoI)- in addition to the background? Since the background statistics depend on unknown parameters (for instance mean and covariance matrix) a set of training samples $\Z \in \mathbb{R}^{p \times n}$, hopefully free of the SoI $\vt$, is observed whose statistics are assumed to match those of the background in the PUT. These training samples are gathered in the vicinity of the PUT (local detection) or along the whole image (global detection). 

Recently in \cite{Vincent20} we addressed sub-pixel detection using the replacement model under a Gaussian background, and we derived the plain generalized likelihood ratio test (GLRT) by maximizing  the joint distribution of $(\vy,\Z)$ with respect to all unknown parameters. Moreover, motivated by some limitations of the replacement model, especially the fact that the filling factor of a sub-pixel target may not be in practice as large as expected, we also derived the GLRT for a mixed model where presence of a target induces a partial replacement of the background \cite{Vincent20b}. These two detectors assume a Gaussian background. However, evidence of non-Gaussianity of hyperspectral data has been brought \cite{Manolakis03,Matteoli10} and therefore it is of interest to extend detectors originally devised for Gaussian background to the broader class of elliptically contoured (EC) distributions \cite{Anderson90,Tai90}.  The aim of this communication  is thus to extend our recent GLRTs from the Gaussian case to the matrix variate $t$-distributed case. We will show that the  GLRTs coincide with their Gaussian counterparts. The paper is organized as follows. In section \ref{section:glrt}, we consider each of the three models and derive the corresponding GLRTs. The latter are evaluated in section \ref{section:numerical} on simulated data but where the target spectral signature, the mean and covariance matrix of the background are obtained from real hyperspectral images.

\section{GLRT for matrix variate $t$-distributed background \label{section:glrt}}
As stated before, let us assume that we wish to decide whether a given vector $\vy$ contains a signature of interest $\vt$ in the presence of disturbance $\vz$ whose mean value $\vmu$ and covariance matrix $\mSigma$ are unknown, and let us assume that a set of training samples $\vz_{i}$, $i=1,\ldots,n$ are available which share the same distribution as $\vz$. These samples can be collected around the PUT or along the whole image. We simply assume here that $n > p$. Therefore, we would like to solve the following problem:
\begin{align}
	H_{0}&: \vy = \vz; \quad \vz_{i} \dist \vz, \, i=1,\ldots,n \nonumber \\
	H_{1} &: \vy = \alpha \vt  + \beta \vz; \quad \vz_{i} \dist \vz, \, i=1,\ldots,n
	\label{prob_glr_generic}
\end{align}
where $\dist$ means ``has the same distribution as''. In \eqref{prob_glr_generic}, $\vt$ corresponds to the assumed spectral signature of the target and $\alpha$ denotes its unknown amplitude. When $\beta=1$ one obtains  the conventional additive model. When  $\beta=1-\alpha$ the replacement model is recovered, and the mixed model corresponds to an arbitrary $\beta$.

In order to derive the GLRT, we need to specify the joint distribution of $\vy$ and $\Z$ where $\Z = \begin{bmatrix} \vz_{1} & \vz_{2} & \ldots & \vz_{n} \end{bmatrix}$. As said in the introduction, we assume that $\yZ$ follows a matrix-variate $t$-distribution with $\nu$ degrees of freedom so that we need to solve the following composite hypothesis testing problem:
\begin{align}
	H_{0}&: \yZ \dist \matrixT{p}{n+1}{\nu}{\M_{0}}{(\nu-2)\mSigma}{\I_{n+1}} \nonumber \\
	H_{1} &: \yZ \dist \matrixT{p}{n+1}{\nu}{\M_{1}}{(\nu-2)\mSigma}{\begin{pmatrix} \beta^{2} & \vect{0}^{T} \\ \vect{0} & \I_{n} \end{pmatrix}}
	\label{prob_glr_generic_student}
\end{align}
where $\M_{0} = \meanyZ$,  $\M_{1} = \begin{bmatrix}\alpha\vt + \beta\vmu & \vmu\vone_{n}^{T} \end{bmatrix}$,  $\vone_{n}$ is a $n\times 1$ vector with all elements equal to one,  $\vmu$ stands for the mean value of the background while $\mSigma$ denotes its covariance matrix.  In \eqref{prob_glr_generic_student}, $\mathcal{T}()$ stands for the matrix variate $t$-distribution \cite{Gupta00,Kotz04}  so that the probability density function (p.d.f.) of the observations under each hypothesis is given by
\begin{align}\label{p0_p1_Student}
	&p_{0}(\vy,\Z)  = C  \det{\mSigma}^{-\frac{n+1}{2}}  \det{\I_{p} + \frac{\invSigma}{\nu-2} \cyZ \cyZtranspose}^{-\frac{\nu+n+p}{2}} \nonumber \\
	& p_{1}(\vy,\Z) =C   \beta^{-p} \det{\mSigma}^{-\frac{n+1}{2}}  \det{\I_{p} + \frac{\invSigma}{\nu-2} \cytildeZ \cytildeZtranspose}^{-\frac{\nu+n+p}{2}}
\end{align}
with $\vytilde = \beta^{-1} (\vy-\alpha\vt)$ and $C = \frac{\Gamma_{p}((\nu+n+p)/2)}{\pi^{p(n+1)/2}\Gamma_{p}((\nu+p-1)/2)}$.  It should be observed that the columns of $\begin{bmatrix} \vy & \Z \end{bmatrix}$ are only uncorrelated but not independent, as $p(\vy,\vz_{1},\ldots,\vz_{n})$ cannot be factored as $p(\vy)\prod_{i=1}^{n}p(\vz_{i})$. 

We now derive the GLRT for the problem in \eqref{prob_glr_generic_student}. Let us start by considering the following function $f(\mSigma)$ where $\S$ is some positive definite matrix:
\begin{align}
	f(\mSigma) &= \det{\mSigma}^{-\frac{n+1}{2}} \det{\I_{p} + (\nu-2)^{-1}\invSigma \S }^{-\frac{\nu+n+p}{2}} \nonumber \\
	&= \det{\mSigma}^{\frac{\nu+p-1}{2}} \det{\mSigma + (\nu-2)^{-1}\S }^{-\frac{\nu+n+p}{2}}
\end{align}
Differentiation of  $\log f(\mSigma) $ yields
\begin{align}
	\frac{\partial \log f(\mSigma)}{\partial \mSigma} = \frac{\nu+p-1}{2} \mSigma^{-1} -\frac{\nu+n+p}{2}(\mSigma + (\nu-2)^{-1}\S )^{-1}.
\end{align}
Setting this derivative of to zero, we can see that $f(\mSigma)$ achieves its maximum at 
\begin{equation}
	\mSigma_{\ast} = \frac{(\nu+p-1)\S}{(\nu-2)(n+1)} = \gamma \S
\end{equation}
It follows that
\begin{align}\label{max_Sigma_student}
	\max_{\mSigma} p_{0}(\vy,\Z) &= C' \det{\cyZ \begin{bmatrix} (\vy - \vmu)^{T} \\ (\Z - \vmu \vone_{n}^{T})^{T} \end{bmatrix}}^{-\frac{n+1}{2}}  \nonumber \\
	\max_{\mSigma} p_{1}(\vy,\Z) &= C' \beta^{-p} \det{\cytildeZ \begin{bmatrix} (\vytilde- \vmu)^{T} \\ (\Z - \vmu \vone_{n}^{T})^{T} \end{bmatrix}}^{-\frac{n+1}{2}}
\end{align}
with $C'=C \gamma^{-p(n+1)/2} \left[1+(\nu-2)^{-1}\gamma^{-1}\right]^{-p(\nu+n+p)/2}$. Since $C'$ is the same under $H_{0}$ and $H_{1}$ it will cancel out in the GLR and therefore the latter does not depend on $\nu$. 
Now, for any vector $\vx$,
\begin{align}
	\M(\vmu) &=\cxZ \cxZtranspose \nonumber \\
	&= (\vx-\vmu)(\vx-\vmu)^{T} + (\Z - \vmu \vone_{n}^{T})(\Z - \vmu \vone_{n}^{T})^{T} \nonumber \\
	&= \vx\vx^{T} - \vmu\vx^{T} - \vx\vmu^{T} + \vmu\vmu^{T} \nonumber \\
	& + \Z\Z^{T} - \vmu\vone_{n}^{T}\Z^{T} - \Z\vone_{n}\vmu^{T} + n \vmu\vmu^{T} \nonumber \\
	&= (n+1)\vmu\vmu^{T} - \vmu(\vx+\Z\vone_{n})^{T} - (\vx+\Z\vone_{n})\vmu^{T} \nonumber \\
	&+ \vx\vx^{T} + \Z\Z^{T} \nonumber \\
	&= (n+1) \left[\vmu -\frac{\vx+\Z\vone_{n}}{n+1}\right] \left[\vmu -\frac{\vx+\Z\vone_{n}}{n+1}\right]^{T} \nonumber \\
	&+ \vx\vx^{T} + \Z\Z^{T}  - \frac{ (\vx+\Z\vone_{n}) (\vx+\Z\vone_{n})^{T}}{n+1} \nonumber \\
	&= (n+1) \left[\vmu -\frac{\vx+\Z\vone_{n}}{n+1}\right]\left[\vmu -\frac{\vx+\Z\vone_{n}}{n+1}\right]^{T} \nonumber \\
	&+ \xZ \left(\I_{n+1} - \frac{\vone_{n+1}\vone_{n+1}^{T}}{n+1}\right) \xZtranspose
\end{align}
Consequently,
\begin{equation}
	\min_{\vmu} \det{\M(\vmu)} = \det{\xZ \Porth{n+1} \xZtranspose}
\end{equation}
with $\Porth{n+1} $ the orthogonal projector on the null space of $\vone_{n+1}$. Hence, we arrive at
\begin{align}
	\max_{\vmu,\mSigma} p_{0}(\vy,\Z) &= C' \det{\yZ \Porth{n+1}\yZtranspose}^{-\frac{n+1}{2}}  \nonumber \\
	\max_{\vmu,\mSigma} p_{1}(\vy,\Z) & =C' \beta^{-p} \det{\ytildeZ \Porth{n+1} \ytildeZ^{T}}^{-\frac{n+1}{2}}
\end{align}
Next, for any vector $\vx$ and matrix $\Q$ (not necessarily $\Porth{n+1} $),
\begin{align}
	\xZ \Q \xZtranspose &= \xZ \begin{bmatrix} Q_{11} & \Q_{12} \\ \Q_{21} & \Q_{22} \end{bmatrix} \xZtranspose \nonumber \\
	&= Q_{11} \vx\vx^{T} + \Z\Q_{21}\vx^{T} + \vx\Q_{12}\Z^{T} + \Z\Q_{22}\Z^{T} \nonumber \\
	&= Q_{11} \left[\vx + Q_{11}^{-1}\Z\Q_{21}\right] \left[\vx + Q_{11}^{-1}\Z\Q_{21}\right]^{T} + Z\Q_{2.1}\Z^{T}
\end{align}
where $\Q_{2.1} = \Q_{22} - \Q_{21}Q_{11}^{-1}\Q_{12}$. Therefore,
\begin{align}
	&\det{\xZ \Q \xZtranspose} = \det{\Z\Q_{2.1}\Z^{T}} \left[1 + Q_{11} \left(\vx + Q_{11}^{-1}\Z\Q_{21}\right)^{T} \left(\Z\Q_{2.1}\Z^{T}\right)^{-1}  \left(\vx + Q_{11}^{-1}\Z\Q_{21}\right)\right]
\end{align}
Coming back to the case $\Q=\Porth{n+1} = \I_{n+1}-(n+1)^{-1}\vone_{n+1}\vone_{n+1}^{T}$, we have
\begin{equation}
	\Q = \begin{pmatrix} 1 & \vect{0}^{T} \\ \vect{0} & \I_{n} \end{pmatrix} - (n+1)^{-1} \begin{pmatrix} 1 & \vone_{n}^{T} \\ \vone_{n} & \vone_{n}\vone_{n}^{T} \end{pmatrix}
\end{equation}
so that
\begin{align}
	Q_{11} &= 1 - (n+1)^{-1} = n(n+1)^{-1} \nonumber \\
	\Q_{21} &= - (n+1)^{-1} \vone_{n} \nonumber \\
	\Q_{22} &=  \I_{n}-(n+1)^{-1}\vone_{n}\vone_{n}^{T} \nonumber \\
	\Q_{2.1} &= \I_{n}-n^{-1}\vone_{n}\vone_{n}^{T}  = \Porth{n}
\end{align}
It follows that $Q_{11}^{-1}\Z\Q_{21} = - n^{-1} \Z\vone_{n} = -\vzbar$ and $\Z\Q_{2.1}\Z^{T} = \Z\Porth{n}\Z^{T} = \Z\Z^{T} - n\vzbar\vzbar^{T} = \S$. Hence, the GLR is given by
\begin{align}\label{GLR_generic}
	\GLR &= \frac{\left[1+Q_{11}(\vy-\vzbar)^{T}\S^{-1}(\vy-\vzbar)\right]^{(n+1)/2}}{\min_{\alpha,\beta}\beta^{p}\left[1+Q_{11}(\vytilde-\vzbar)^{T}\S^{-1}(\vytilde-\vzbar)\right]^{(n+1)/2}} \nonumber \\
	&=\frac{[1+\frac{n}{n+1}(\vy-\vzbar)^{T}\S^{-1}(\vy-\vzbar)]^{(n+1)/2}}{\min_{\alpha,\beta}\beta^{p}[1+\frac{n}{n+1}(\frac{\vy-\alpha\vt}{\beta}-\vzbar)^{T}\S^{-1}(\frac{\vy-\alpha\vt}{\beta}-\vzbar)]^{(n+1)/2}} 
\end{align}

A few important observations can be made regarding this result. First, for all three models, \emph{the GLRs in \eqref{GLR_generic} coincide with their Gaussian counterparts}. For the additive model a proof is given in Appendix \ref{app:conventional_Gaussian}.  A more intuitive way to figure out this equivalence is to realize that the expression of the GLR in \eqref{GLR_generic} does not depend on $\nu$ and that, letting $\nu$ grow to infinity, one should recover the GLR for Gaussian distributed data. As for the replacement and the mixed models, the expression in \eqref{GLR_generic} is exactly that of the ACUTE and SPADE detectors of \cite{Vincent20} and \cite{Vincent20b} respectively, where the GLRTs for the replacement model and the mixed model are derived under the Gaussian assumption. Therefore, the latter are still GLRTs for a much broader class of distributions than initially expected. 

Let us also briefly comment on the implementation of the GLRT. For the additive model $\beta=1$, and the minimization problem in \eqref{GLR_generic} is a simple linear least-squares problem for which a closed-form solution can be obtained. This yields
\begin{align}\label{GLR_additive}
	\GLRam^{2/(n+1)} &= \frac{1+\frac{n}{n+1}(\vy-\vzbar)^{T}\S^{-1}(\vy-\vzbar)}{\min_{\alpha}1+\frac{n}{n+1}(\vy-\vzbar-\alpha\vt)^{T}\S^{-1}(\vy-\vzbar-\alpha\vt)} \nonumber \\
	&= \frac{1+\frac{n}{n+1}(\vy-\vzbar)^{T}\S^{-1}(\vy-\vzbar)}{1+\frac{n}{n+1}(\vy-\vzbar)^{T}\S^{-1}(\vy-\vzbar) - \frac{n}{n+1}\frac{[(\vy-\vzbar)^{T}\S^{-1}\vt]^{2}}{\vt^{T}\S^{-1}\vt}} \nonumber \\
	&\equiv \frac{\frac{n}{n+1}[(\vy-\vzbar)^{T}\S^{-1}\vt]^{2}}{[1+\frac{n}{n+1}(\vy-\vzbar)^{T}\S^{-1}(\vy-\vzbar)][\vt^{T}\S^{-1}\vt]}
\end{align}
The GLR in \eqref{GLR_additive} generalizes Kelly's detector to the case of a non-centered Student distributed background. Note that \eqref{GLR_additive} differs  from Kelly's detector by the $\frac{n}{n+1}$ factor.

As for the replacement model, $\beta=1-\alpha$ and the minimization should be conducted with respect to $\alpha$ only, i.e.,
\begin{align}\label{ACUTE}
	\GLRrm = &\frac{\left[1+\frac{n}{n+1}\left\|\S^{-1/2}(\vy-\vzbar)\right\|^{2}\right]^{(n+1)/2}}{\min_{\alpha}(1-\alpha)^{p}\left[1+\frac{n}{n+1}\frac{\left\|\S^{-1/2}(\vy-\alpha\vt - (1-\alpha)\vzbar)\right\|^{2}}{(1-\alpha)^{2}}\right]^{(n+1)/2}} 
\end{align}
As shown in \cite{Vincent20},  this simply amounts to finding the (unique) positive root of a 2nd-order polynomial. Finally, for the mixed model where $\beta$ is arbitrary, one has a 2-D minimization problem. However, minimization over $\alpha$ can be done analytically, leaving only a minimization over $\beta$:
\begin{align}\label{SPADE}
	\GLRmm =\frac{\left[1+\frac{n}{n+1}\left\|\S^{-1/2}(\vy-\vzbar)\right\|^{2}\right]^{(n+1)/2}}{\min_{\beta}\beta^{p}\left[1+\frac{n}{n+1}\frac{\left\|\Porth{\S^{-1/2}\vt}\S^{-1/2}(\vy-\beta\vzbar)\right\|^{2}}{\beta^{2}}\right]^{(n+1)/2}} 
\end{align}
Again the solution is obtained as the unique positive root of a second-order polynomial equation \cite{Vincent20b}. Therefore for both the replacement model and the mixed model, all unknown parameters can be obtained in closed-form.

\section{Numerical simulations \label{section:numerical}}
In the present section, we will compare the detectors developed above. The GLR \eqref{GLR_additive}  will be referred to as Kelly in the sequel as it generalizes the original GLRT of Kelly \cite{Kelly86} to a non-centered Student distributed background. The GLRs in \eqref{ACUTE} and \eqref{SPADE} will be referred as to ACUTE  \cite{Vincent20} and SPADE \cite{Vincent20b}. These last two detectors have already been assessed against real data drawn from the RIT and Viareggio experiments. Herein, we evaluate their performance as well as their robustness on simulated yet realistic data. More precisely, we consider $\vt$ to be the signature of the V5 target in the Viareggio image \cite{Acito16}, $\vmu$ and $\mSigma$ are respectively the sample mean and sample covariance matrix obtained from the whole Viareggio image. It has to be noticed that these raw radiance data have first been converted to reflectance measurements using an ELM method \cite{Ferrier95,Smith99}. The number of spectral bands used is $p=32$ and the number of training samples is $n=60$. The background is simulated using a $t$ distribution with $\nu=5$ degrees of freedom.

\begin{figure}[htb]
	\centering
	\includegraphics[width=10cm]{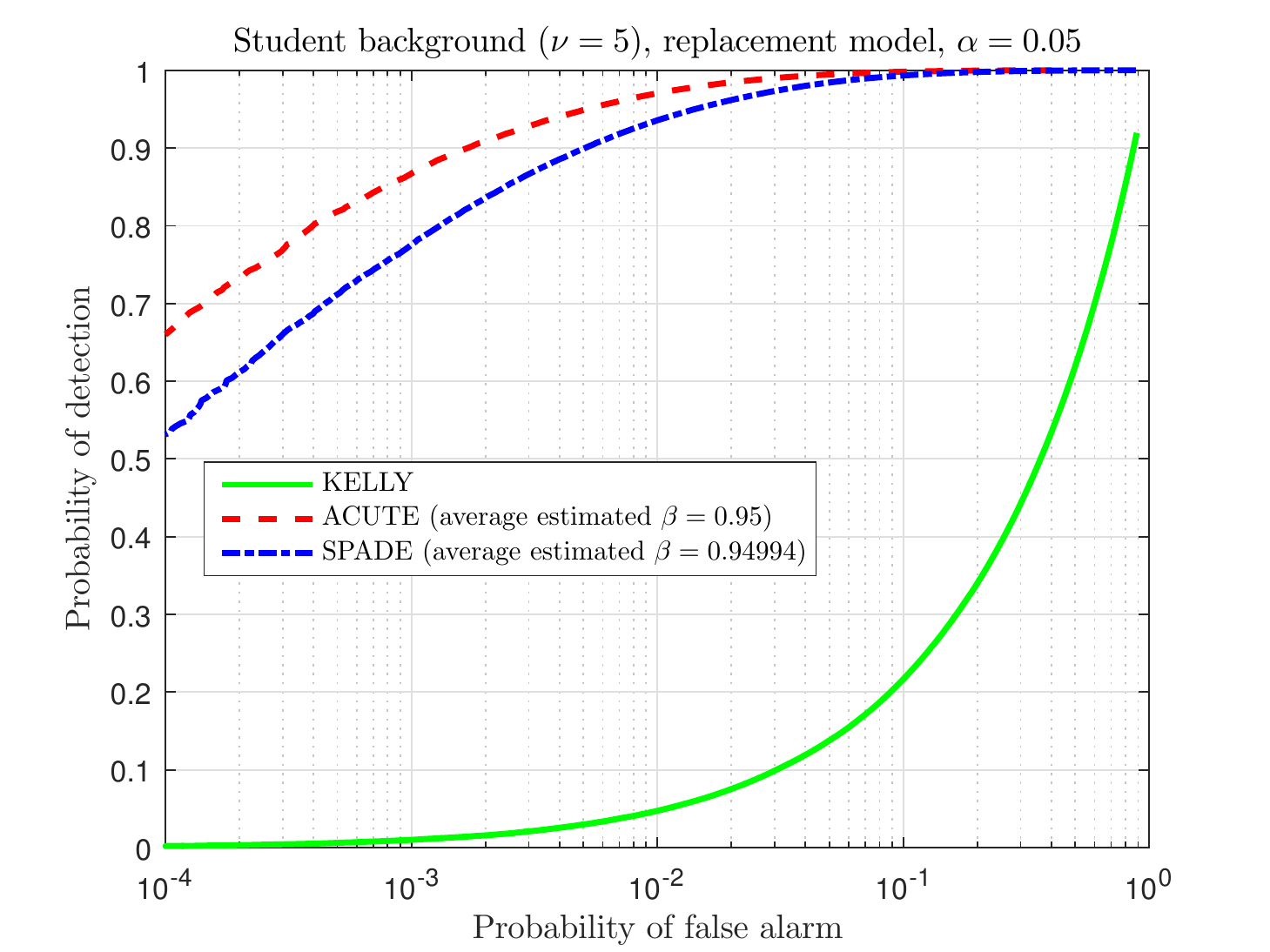} 
	\caption{ROC for the replacement model $\beta=(1-\alpha)$.}
	\label{fig:cor}
\end{figure}

Figure \ref{fig:cor} plots the Receiver Operation Curve (ROC) obtained for the replacement model [$\beta=1-\alpha$] with $\alpha=0.05$. As could be anticipated, ACUTE exhibits the best performance since  it corresponds to the GLRT for the specific case $\beta=1-\alpha$. However, SPADE is shown to incur a small degradation compared to this optimal detector. On the contrary, Kelly exhibits a significant performance loss, mostly because $\E{\vz} \neq \E{\vz_{k}}$,  a fact that is not accounted for in the additive model, contrary to the other two detectors.\\

\begin{figure}[htb]
	\centering
	\includegraphics[width=10cm]{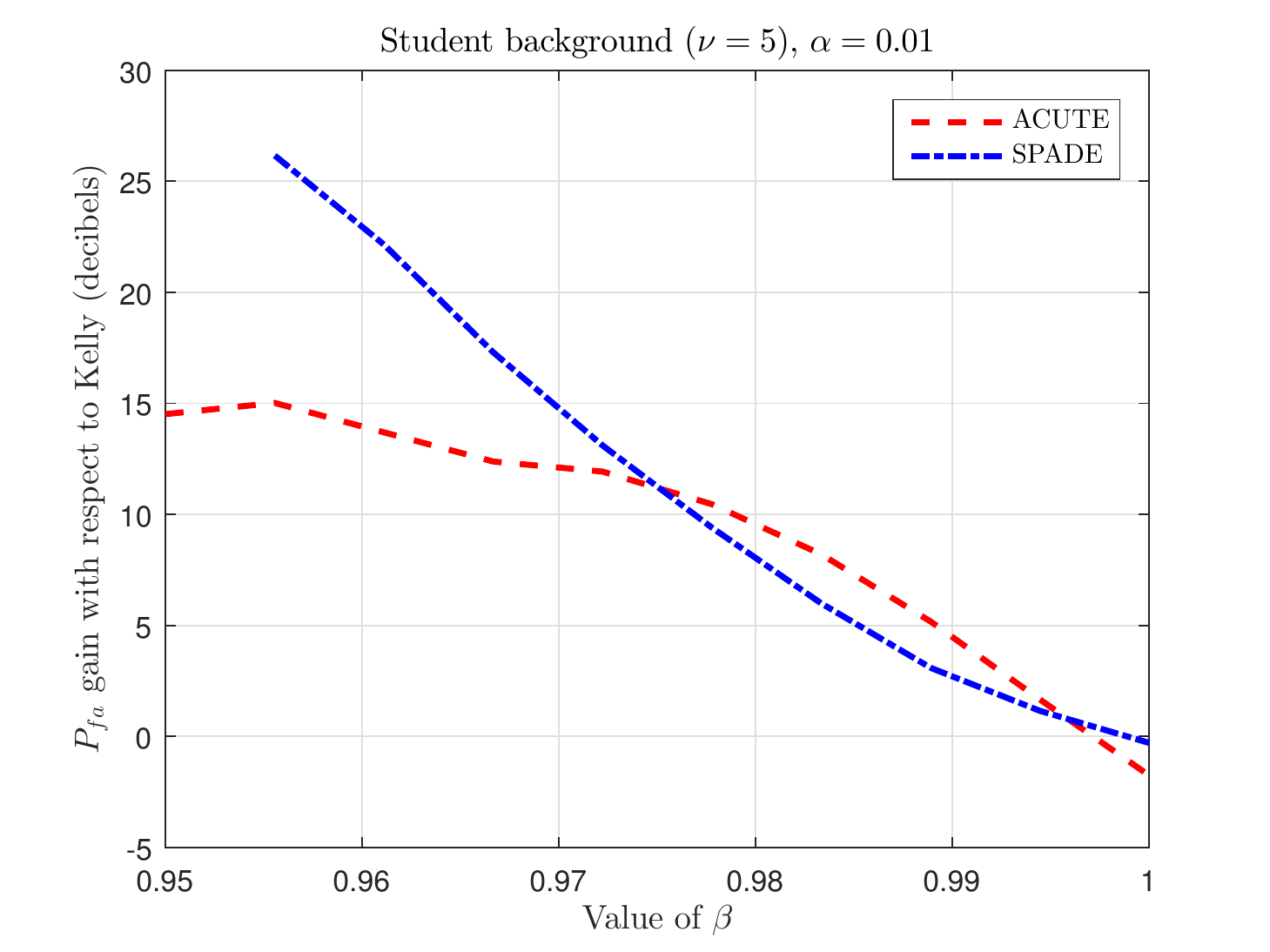} 
	\caption{$\Pfa$ gain versus $\beta$. $\alpha=0.01$ and $\Pd=0.5$.}
	\label{fig:Pfa_gain}
\end{figure}

We now assess the robustness of ACUTE and SPADE. More precisely, we study their performance when $\beta$ varies. In Figure \ref{fig:Pfa_gain}, we display the probability of false alarm ($\Pfa$) gain of ACUTE and SPADE with respect to Kelly, i.e., $10\log_{10}\frac{\Pfa(\GLRam)}{\Pfa(\GLR_{\text{\tiny{RM/MM}}})}$.  In this figure  $\alpha$ is fixed to $0.01$  and the probability of detection is $\Pd=0.5$. Figure \ref{fig:Pfa_gain} confirms that ACUTE is slightly better than SPADE when the acual value of $\beta$ is close to $1-\alpha$. However, as soon as $\beta$ departs from $1-\alpha$ SPADE shows better performance. Moreover, SPADE does not incur any loss compared to Kelly when $\beta=1$, contrary to ACUTE. Therefore, SPADE provides the best robustness, with small performance loss compared to the optimal solution whatever the value of $\beta$.

\section{Concluding remarks}
In this communication, we considered the detection of a sub-pixel target in hyperspectral imaging when the background is no longer Gaussian but $t$-distributed. The GLRTs for a general mixed model, including the standard additive case and the replacement one,  were derived, generalizing the Gaussian versions previously derived. For the three specific values of $\beta$ considered in the literature, it was shown that the GLRTs remain the same and hence the detectors initially proposed under a Gaussian framework have more generality than expected. Moreover, they do not depend on the unknown degree of freedom of the $t$-distribution. Numerical simulations showed that SPADE provides a very good trade-off as it is always close to or better than Kelly and ACUTE which are optimal only for specific values of $\beta$.

\appendix 
\section{GLR for the additive model and Gaussian distributed background\label{app:conventional_Gaussian}}
In this appendix, we derive the GLRT for Gaussian distributed background and for the additive model. We thus consider the following detection problem
\begin{align}
	H_{0}&: \yZ \dist \matrixN{p}{n+1}{\meanyZ}{\mSigma}{\I_{n+1}} \nonumber \\
	H_{1} &: \yZ \dist \matrixN{p}{n+1}{\meantyZ}{\mSigma}{\I_{n+1}}
\end{align}
The p.d.f. of $(\vy,\Z)$ is in this case
\begin{align}
	&p_{0}(\vy,\Z) = (2\pi)^{-\frac{(n+1)p}{2}}  \det{\mSigma}^{-\frac{n+1}{2}} \etr{-\frac{1}{2}  \invSigma \cyZ \begin{bmatrix} (\vy - \vmu)^{T} \\ (\Z - \vmu \vone_{n}^{T})^{T} \end{bmatrix}}\nonumber \\
	&p_{1}(\vy,\Z) = (2\pi)^{-\frac{(n+1)p}{2}}   \det{\mSigma}^{-\frac{n+1}{2}} \etr{-\frac{1}{2}  \invSigma \cyZ \begin{bmatrix} (\vy - \vmu)^{T} \\ (\Z - \vmu \vone_{n}^{T})^{T} \end{bmatrix}}
\end{align}
It is well-known that $\det{\mSigma}^{-\frac{n+1}{2}} \etr{-\frac{1}{2}  \invSigma\S }$ achieves its maximum at $\mSigma_{\ast} = (n+1)^{-1}\S$, and hence
\begin{equation}
	\max_{\mSigma} \det{\mSigma}^{-\frac{n+1}{2}} \etr{-\frac{1}{2}  \invSigma\S } = \left(\frac{e}{n+1}\right)^{-\frac{n+1}{2}} \det{\S}^{-\frac{n+1}{2}} 
\end{equation} 
It follows that $\max_{\mSigma} p_{0}(\vy,\Z)$ and $\max_{\mSigma} p_{1}(\vy,\Z)$ are proportional to \eqref{max_Sigma_student} which holds for Student distributions. From there, everything follows and the GLRs for Student or Gaussian distributions are the same and are given by \eqref{GLR_generic}.

\end{document}